\documentclass[aps,prl,reprint,superscriptaddress]{revtex4-1}
\usepackage{graphicx}

\begin{document}

\title{Thermally driven anomalous Hall effect transitions in FeRh}

\author{Adrian Popescu}
\affiliation{Department of Physics, University of South Florida, Tampa, Florida 33620, USA}

\author{Pablo Rodriguez-Lopez}
\affiliation{Materials Science Factory, Instituto de Ciencia de Materiales de Madrid, ICMM-CSIC, Cantoblanco, E-
28049 Madrid, Spain}
\affiliation{GISC-Grupo Interdisciplinar de Sistemas Complejos, 28040 Madrid, Spain}

\author{Paul M. Haney}
\affiliation{Center for Nanoscale Science and Technology, National Institute of Standards and Technology, Gaithersburg, Maryland 20899, USA}

\author{Lilia M. Woods}
\affiliation{Department of Physics, University of South Florida, Tampa, Florida 33620, USA}

\date{\today}

\begin{abstract}
Materials exhibiting controllable magnetic phase transitions are currently in demand for many spintronics applications. Here we investigate from first principles the electronic structure and intrinsic anomalous Hall, spin Hall and anomalous Nernst response properties of the FeRh metallic alloy which undergoes a thermally driven antiferromagnetic-to-ferromagnetic phase transition.  We show that the energy band structures and underlying Berry curvatures have important signatures in the various Hall effects. Specifically, the suppression of the anomalous Hall and Nernst effects in the AFM state and a sign change in the spin Hall conductivity across the transition are found. It is suggested that the FeRh can be used a spin current detector capable of differentiating the spin Hall effect from other anomalous transverse effects. The implications of this material and its thermally driven phases as a spin current detection scheme are also discussed.  
\end{abstract}


\maketitle

Hall effect phenomena are currently being studied in the context of new materials, including 2D systems, topological insulators, and Weyl materials among others \cite{Ezawa2013,Sinova2015,Nagaosa2010}. Fundamental discoveries of novel 
properties  and their applications beyond the standard Hall effect are an important step toward new devices \cite{Sinova2015,Nagaosa2010}. However, even in typical materials the current understanding of these phenomena is not complete. The standard Hall effect is present in all conductors under an external magnetic field. Materials with broken time reversal symmetry, such as ferromagnets (FMs), can exhibit the anomalous Hall effect (AHE) and materials with strong spin orbit coupling (SOC), such as heavy metals, can exhibit the spin Hall effect (SHE). Although the AHE and SHE have intrinsic and extrinsic origins \cite{Luttinger1954,Smit1955,Berger1970}, the occurrence of these effects has been mostly associated with extrinsic contributions due to spin-dependent skew and side-jump scattering mechanisms \cite{Gradhand2010,Sinitsyn2007}. Recent reports, however, have shown that the intrinsic contribution, stemming from the materials electronic structure and its Berry phase features, is the primary source for the AHE and SHE in many systems \cite{MacDonald2002,Niu2004,Nagaosa2008}. 

Further developments strongly support that Hall effects in metals with moderate conductivity are intrinsic in nature \cite{Sinova2015,Nagaosa2010}. These results are important for spintronics applications, which rely on the generation and detection of spin currents based on the  SHE and the inverse spin Hall effect (ISHE) \cite{Sinova2015}. Spin currents can deliver large amounts of angular momentum with minimum power dissipation and Joule heating, which makes them attractive for technology. Their detection and generation are not trivial and require readily accessible electric components, thus new interpretations of SHE and ISHE in materials with strong SOC are highly desirable.

Typical materials in spintronics are FM and nonmagnetic metals, although recently noncollinear antiferromagnets (AFMs) and Weyl semimetals have been shown to exhibit large intrinsic AHE and SHE \cite{Yan2017,Sun2016}. Investigating intrinsic Hall effects in the context of particular lattices \cite{Chudnovsky2009} may hold promise for bringing fundamental science forward and resolving technological challenges. Other systems that can give a different perspective of various Hall effects are materials that exhibit AFM-FM transitions. In particular,  we suggest that the temperature driven magnetic phase transition of the equiatomic ordered metallic alloy FeRh has important consequences for the spin phenomena supported by this material. 

FeRh has a CsCl-like crystal structure \cite{Kubaschewski1982,Fallot1939,Kouvel1962} and it undergoes an AFM to FM transition at $\approx$ 350 K, which makes this material interesting for fundamental studies and devices, including for magnetic recording, media storage, and magnetocaloric cycles \cite{Fullerton2003,Ramesh2014,Dkhil2016}. FeRh is also attractive for spin transport applications due to its large SOC from the Rh atom. Surprisingly, a limited number of studies have been reported on Hall measurements mostly utilizing external magnetic fields and doping \cite{Kobayashi2001,Marrows2013}, which strongly suggests that this area is largely unexplored.

\begin{figure*}
\includegraphics[width=\textwidth,height=7cm]{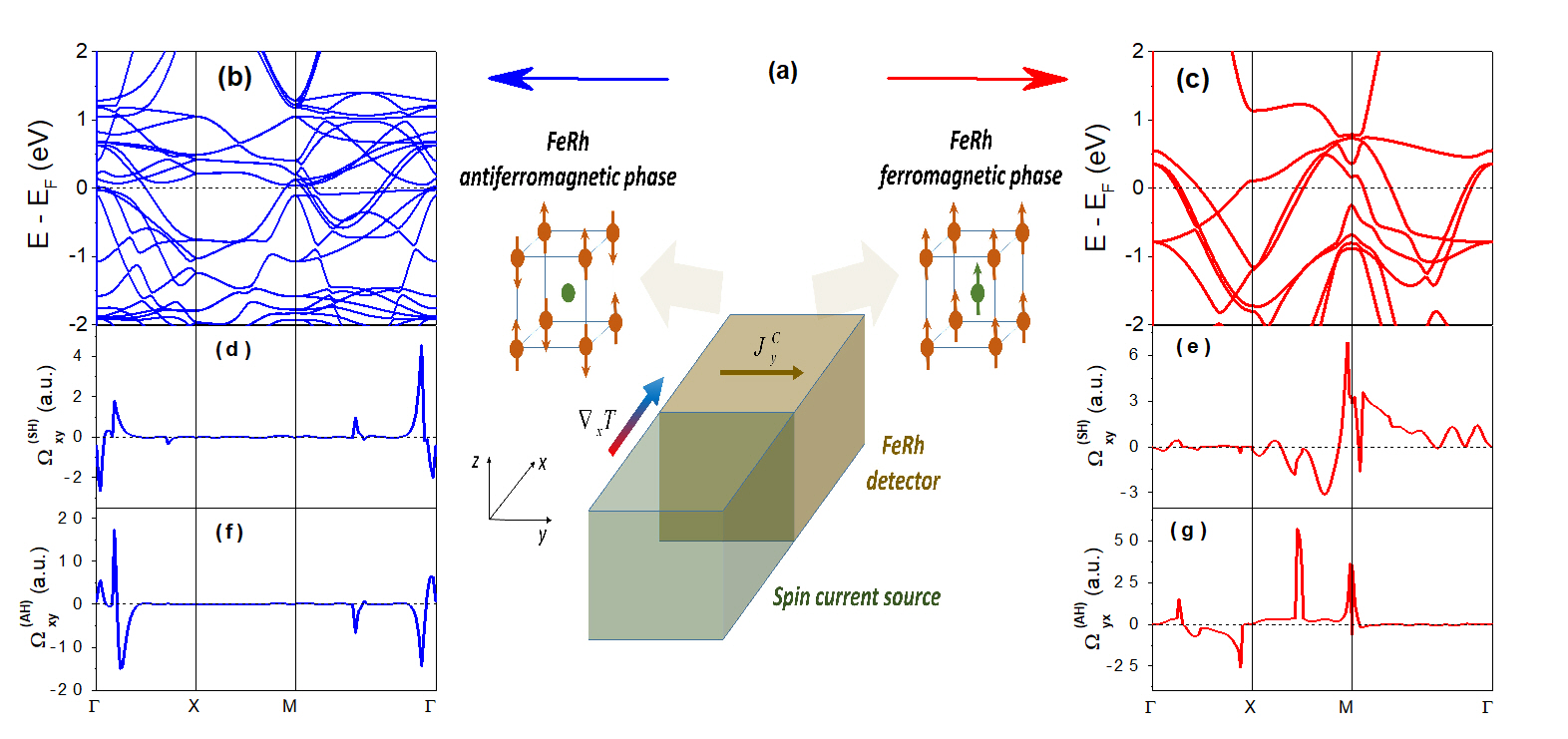}
\caption{\label{figure1} (Color online) (a) A spin current detection set-up: a spin current is thermally injected along the $x$-axis into a FeRh detector, in which a transverse charge current in the $y$-direction, $J_y^C$ is generated. (b), (c) Energy band structures for the AFM and FM magnetic phases, respectively. (d-g) Berry curvature and spin Berry curvature for both magnetic phases, summed over all occupied bands, and calculated along the same $k$-space symmetry lines. The FeRh latticess and spin polarizations for both phases are also shown.}
\end{figure*}

In this work, we investigate the intrinsic Hall effect phenomena in the AFM and FM phases of FeRh from first principles. Several properties, including the spin Hall, anomalous Hall, and  anomalous Nernst conductivities, originating from the band structure and geometry through their Berry curvatures, are calculated based on the Kubo linear response formalism \cite{Mahan2000}. We demonstrate that the thermally controlled AFM-FM transition leads to important modifications in the FeRh transverse Hall effect responses. We further suggest that this material may be suitable for spin current detection in a longitudinal configuration, where a temperature gradient is used to drive a spin current from a ferromagnetic source. In current practice \cite{Holanda2017,Miao2016,Uchida2014}, the AHE and the anomalous Nernst effect (ANE) can result into additional unwanted contributions to the transverse charge current, which must be separated from the contribution associated with the longitudinal spin current. Our results below show that the thermally driven changes in the FeRh  Hall response may be useful for mitigating measurement artifacts arising from AHE and ANE. 

For the spin current measurement one can utilize an experimental setup, shown in Fig.  \ref{figure1} (a), where an applied temperature gradient $\nabla_x T$ injects a spin current from the source into a metallic detector made of FeRh, and an electronically measurable charge current is generated in the transverse $y$-direction. Since the magnetic phases of FeRh are metallic, it is expected that $\nabla_x T$ will also induce a longitudinal electric field $E_x$ via the Seebeck effect, which may give rise to an additional contribution to the transverse signal via the AHE. Here we consider the general form of the transverse charge current, which is given as $J_y^C = \sigma_{yx}^{(SH)} \nabla_x \mu_s + \sigma_{yx}^{(AH)} E_x + \sigma_{yx}^{(AN)} (- \nabla_x T)$, where $\mu_s$ is the spin chemical potential and $\sigma_{yx}$ are the conductivities with superscripts for the SHE, AHE, and ANE, respectively. Below we show how the FeRh phases affect the various transverse Hall conductivities.

We first describe the electronic structure properties of both FeRh phases using $ab$ $initio$ simulations. The local density approximation to the density functional theory with SOC included, as implemented in the Quantum ESPRESSO package \cite{Giannozzi2009}, is used throughout the calculations. The ground-state for each phase is obtained by setting the cutoff kinetic energy for the wavefunctions to 180 Ry ($\approx 2.5\times 10^3\, {\rm eV}$), and by using a $20 \times 20 \times 20$ uniform $k$-points grid. Fully relativistic, norm-conserving atomic pseudopotentials are built by using the Atomic Pseudopotential Engine package \cite{ape}, such that the relaxed structures for both phases to match the experimental lattice constants \cite{Lee1970} to less than 0.001 \%.

The Hall conductivities for the ISHE, AHE, and ANE are expressed as:
\begin{equation}
\sigma_{yx}^{(E)} = \sigma_{0}^{(E)} \sum_{n} \int_{BZ} d \mathbf{k} F_{n \mathbf{k}}^{(E)} \Omega_{n,yx}^{(E)} (\mathbf{k}) \label{sigma}
\end{equation}
where the $n$-summation runs over all bands, $E=\{SH, AH, AN\}$  and $\sigma_{0}^{(SH)} = e/(2 \pi)^3$, $\sigma_{0}^{(AH)} = e^2/ \hbar (2 \pi)^3$, $\sigma_{0}^{(AN)} = e/[T \hbar (2 \pi)^3]$. Also, $F_{n \mathbf{k}}^{(SH,AH)} = 1/(1+e^{ (\epsilon_{n \mathbf{k}} - E_{F}) / k_B T})$ represent the Fermi-Dirac distribution and $F_{n \mathbf{k}}^{(AN)} = (\epsilon_{n\mathbf{k}} - E_{F}) F_{n \mathbf{k}}^{(AH)} + k_B T \text{ln} (1+e^{- (\epsilon_{n \mathbf{k}} - E_{F}) / k_B T})$ is the entropy density \cite{Jauho2015,Tewari2012}. The Berry curvature of band $n$ at point $\mathbf{k}$ in the Brillouin zone is given by \cite{Nagaosa2010}:

\begin{equation}
\Omega_{n,yx}^{(AH,AN)}  = -2 \text{Im} \sum_{m \neq n} \frac{\langle u_{n \mathbf{k}} | \hat{v}_x | u_{m \mathbf{k}} \rangle  \langle u_{m \mathbf{k}} | \hat{v}_y | u_{n \mathbf{k}} \rangle}{(\epsilon_{n \mathbf{k}} - \epsilon_{m \mathbf{k}})^2} \label{omega_ah}
\end{equation}
where $\epsilon_{n \mathbf{k}}$ and $| u_{n \mathbf{k}} \rangle$ are the eigenvalues and eigenvectors of the Bloch Hamiltonian $\hat{H}$ and $\hat{v}_i= \hbar^{-1} d \hat{H} / d k_i$ is the velocity operator of the $i$-direction, respectively. The spin Berry curvature is \cite{Sinova2015}:

\begin{equation}
\Omega_{n,yx}^{(SH)}  = -2 \text{Im} \sum_{m \neq n} \frac{\langle u_{n \mathbf{k}} | \hat{J}_x^z | u_{m \mathbf{k}} \rangle  \langle u_{m \mathbf{k}} | \hat{v}_y | u_{n \mathbf{k}} \rangle}{(\epsilon_{n \mathbf{k}} - \epsilon_{m \mathbf{k}})^2} \label{omega_sh}
\end{equation}
where the spin current operator $\hat{J}_x^z = \{ \hat{v}_x , \hat{\sigma}_z \} /2$ describes the spin flow with $z$-polarization along the $x$-direction ($\hat{\sigma}_z$ is the Pauli spin operator).

The calculation of these properties is facilitated by employing the Wannier interpolation technique \cite{Vanderbilt2006}, where the \textit{ab initio} wave functions for each phase are projected on 18 maximally-localized Wannier functions per atom with the Wannier90 code \cite{wannier}. The integration over the occupied states of the Brillouin zone for the Berry curvatures is performed by using an initial $100 \times 100 \times 100$ $k$-point uniform mesh, followed by an adaptively-refined $k$-mesh until a 5 \% convergence is achieved. The electronic band structure for both FeRh phases is shown in Figs. \ref{figure1} (b) and (c) and it corresponds to previously reported density of states (not shown graphically) \cite{Turek2015,Ebert2016}. It has been suggested that the magnetic moments of the Rh atoms play a conclusive role in the AFM-FM transition, although the complete picture of this transition is not fully established yet \cite{Staunton2014}. Our calculations show that the magnetic moments are 2.98 $\mu_B$ and 3.04 $\mu_B$ for the Fe atoms in the AFM and FM states, respectively, and 
1.18 $\mu_B$ for the Rh atoms in FM phase, which is in excellent agreement with previous calculations \cite{Turek2015,Ebert2016} and experiments \cite{Bertaut1962,Stamm2008}.

To understand the role of the Berry and spin Berry curvatures in the conductivities,  in Figs. \ref{figure1} (d-g) we show the summed over all occupied bands $\Omega_{yx}^{(SH,AH)} = \sum_n \Omega_{n,yx}^{(SH,AH)}$ along the energy bands symmetry lines. Both $\Omega_{yx}^{(AH)}$ and $\Omega_{yx}^{(SH)}$ have distinct peak-like structures, with large positive and negative contributions at small regions of the $k$-space. These peaks usually occur when the Fermi level lies between pairs of occupied-unoccupied bands coupled through SOC with small energy separation. For example, as shown in Figs. \ref{figure1} (d) and (f), the peaks near the $\Gamma$-point for the AFM phase, and the peaks along the $X-M$ line for the FM phase in Figs. 1 (e) and (g), are due to such pairs of bands, where the small energy difference gives rise to small denominators in Eqs. (2) and (3). Peak-like structure of $\Omega_{yx}^{(SH,AH)}$ of the same origin is also found in other regions of the Brillouin zone for both phases.

\begin{figure}
\includegraphics[scale=1]{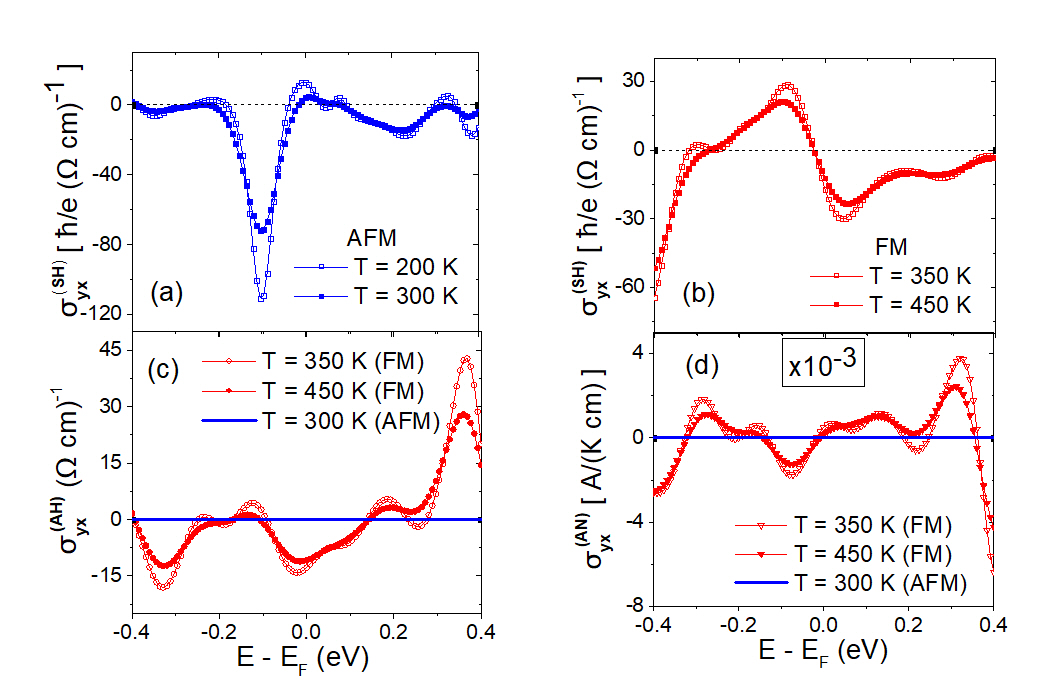}
\caption{\label{figure2} (Color online) (a) Spin Hall conductivity for the AFM state of FeRh, (b) Spin Hall conductivity for the FM state of FeRh, (c) Anomalous Hall conductivity, and (d) Anomalous Nernst conductivity for both FeRh phases. Results for two temperatures are shown.}
\end{figure}

\begin{figure*}
\includegraphics[width=\textwidth,height=8cm]{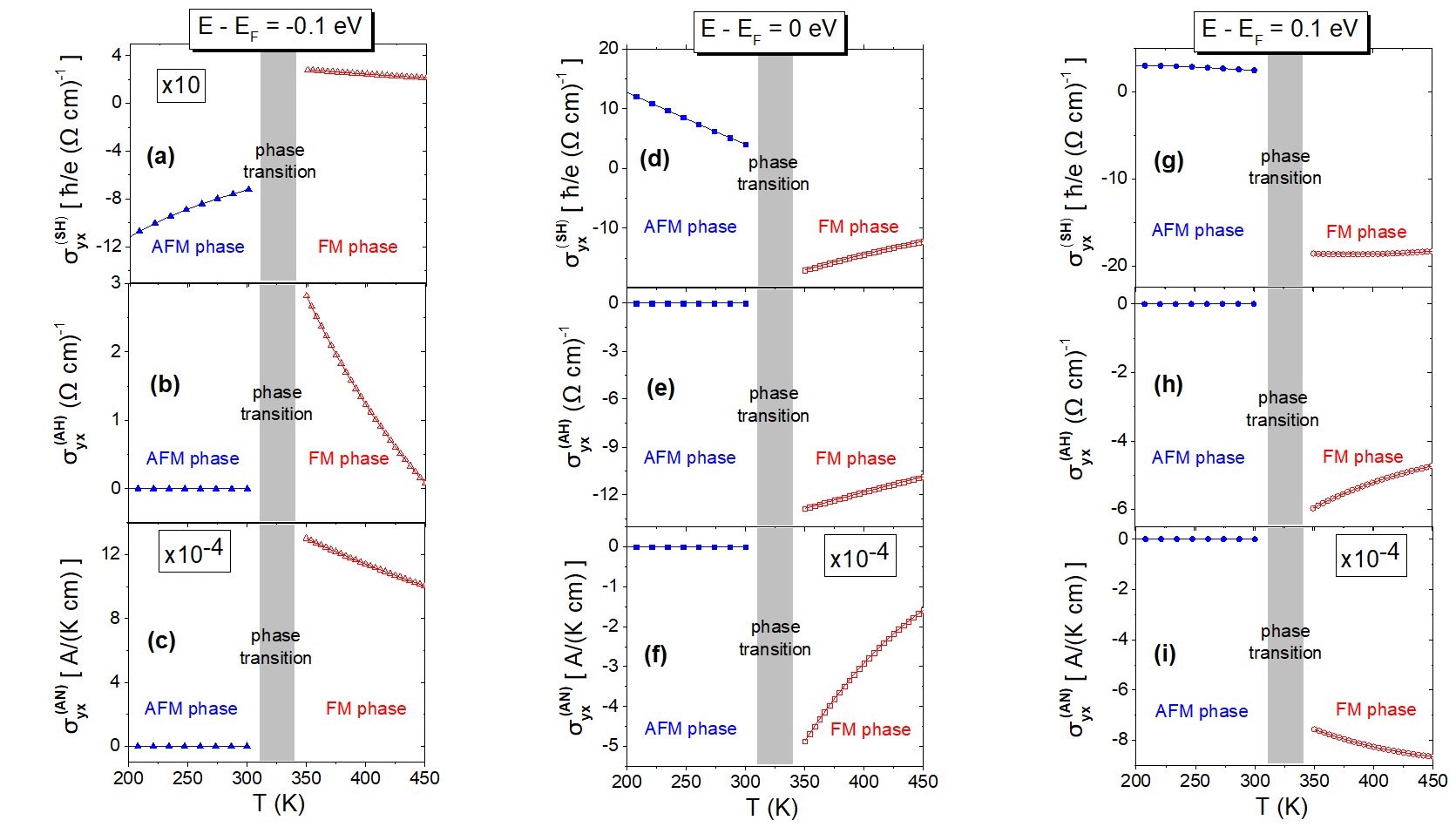}
\caption{\label{figure3} (Color online) Temperature dependence for: (a), (d) and (g) the spin Hall conductivity; (b), (e) and (h) the anomalous Hall conductivity; (f), (i) and (c) the anomalous Nernst conductivity for different values of the Fermi level. The gray region represents the temperature interval where the phase transition occurs.}
\end{figure*}

Results for the various intrinsic conductivities as a function of the Fermi level are shown in Fig. \ref{figure2} for both FeRh  magnetic phases. Similar to $\Omega_{yx}^{(SH,AH)}$, the response is sensitive to the $E_F$ position in the band structure. More importantly, significant differences between the AFM and FM phases are found.  For example, $\sigma_{yx}^{(SH)}$ for the AFM phase exhibits a relatively large negative peak at $E - E_F \approx -0.2$ eV with values comparable to other metallic AFMs found to support large spin currents with intrinsic origin \cite{Mokrousov2014}. Changing  $E_F$ probes different parts of the energy band structure, which induces modifications in $\sigma_{yx}^{(SH)}$ including sign change and different peak locations. The spin Hall conductivity for the FM phase exhibits different behavior as a function of $E_F$, with a pronounced sign switch of $\sigma_{yx}^{(SH)}$ at $E - E_F \approx -0.2$ eV (Fig. \ref{figure2} (b)). We find that although higher $T$ tends to reduce the absolute values of the conductivities and smooth out the peaks, the peak structure is fairly robust. These Berry curvature effects are also important for $\sigma_{yx}^{(AH)}$ and $\sigma_{yx}^{(AN)}$, for which $T$ is found to have a similar role as for the spin Hall conductivity for both phases. The AFM $\sigma_{yx}^{(AH)}$ and $\sigma_{yx}^{(AN)}$ are negligible (shown by the blue horizontal lines in Figs. \ref{figure2} (c) and (d)). This is a direct consequence of the preserved time reversal symmetry of the AFM phase, leading to a vanishing integrated Berry curvature. 

The dramatic changes in the response properties across the phase transition suggest to further explore how $\sigma_{yx}^{(SH)}$, $\sigma_{yx}^{(AH)}$, and $\sigma_{yx}^{(AN)}$ evolve as a function of temperature, which can further be used to analyze the behavior of the induced charge current in FeRh. Our results for the $T$ dependence of the properties for different $E_F$ are given in Fig. \ref{figure3}, which clearly shows the significant changes of all studied conductivities across the phase transition. One notes that $\sigma_{yx}^{(SH)}$ for the AFM is rather large, has a nontrivial behavior as a function of $T$, and can be positive or negative depending on $E_F$ (Figs. \ref{figure3} (a), (d) and (g)). As the material is driven to its FM phase, the corresponding $\sigma_{yx}^{(SH)}$ switches its sign and it evolves almost linearly with temperature. At the same time, while $\sigma_{yx}^{(AH)}$ and $\sigma_{yx}^{(AN)}$ are of negligible magnitude (practically zero) in the AFM phase, their FM counterparts show similar $T$-dependence as $\sigma_{yx}^{(SH)}$.

These thermally driven changes in the intrinsic Hall effect phenomena can be utilized as a reliable spin current detection scheme (shown in Fig. \ref{figure1} (a)). A spin current is injected from the source material into the FeRh detector by applying $\nabla_x T$, followed by measuring the transverse charge current $J_y^C$. Since large $\nabla_x T$ is needed to generate the spin current, it is expected that the FeRh is in its FM state. By thermally driving the FM FeRh into its AFM state one expects the measured transverse signal to change discontinously. If the measured current remains nonzero, then this is an unambiguous signature of the ISHE meaning that a spin current is injected into the FeRh detector. If, however, $J_y^C$ is found to vanish in the AFM phase, then the transverse charge current in the FM phase is entirely the result of the AHE and ANE in FeRh, and there is no longitudinal spin current generated by $\nabla_x T$. Calculations performed at different values of $E_F$ show the same overall behavior (Figs. \ref{figure3} (a-c) and (g-i)), indicating that this detection scheme is also fairly robust against changes in the chemical potential. Direct comparison between our computational results with experimental data  is not possible at this stage due to the very limited measurements of this material across the phase transition (the available experiments are at non-zero magnetic fields and doped samples \cite{Marrows2013}). Nevertheless, typical values for the FeRh electrical conductivity are $\sigma< 10^6$ $\Omega\, $cm \cite{Marrows2013}, which is in the range where skew and side-jump scattering are expected to be not significant and the Hall effect is of intrinsic nature \cite{Sinova2015,Nagaosa2010}. The calculated $\sigma_{yx}^{(SH)}$ values are comparable with those of other materials with significant ISHE and are accessible with current lab capabilities \cite{Mokrousov2014,Behnia2017}. 

Let us further note that at present, most spin current detection set-ups use heavy metals such as Pt, W or Ta \cite{Holanda2017,Miao2016,Uchida2014}, in which the ISHE generates a measurable charge current in a transverse direction via a strong SOC. However, magnetic proximity effects induced in the detector give rise to contaminating transverse charge currents through the AHE and ANE. In practice, these unwanted contributions are usually removed by inserting additional layers between the spin source and the detector \cite{Miao2016,Tian2015}, which is problematic for measurements control and reliability. There is strong evidence, however, that Fe-terminated surfaces in various FeRh/substrates retain the bulk FeRh properties and there are no interface magnetic layers \cite{Shick2015,Kao2010,Alouani2017,comment}, although this issue has to be investigated further with magnetic substrates. By adjusting $T$, the set-up in Fig. \ref{figure1} (a) may be used to resolve the origin of injected spin current, without the need for additional interface layers. In the presence of a proximately magnetized interface layer on the FeRh bulk AFM phase, one expects that the resulting unwanted current is relatively small since it scales as the ratio between the number of proximately magnetized atomic layers (typically $<$5) and the much bigger contact measurement length \cite{Sinova2015,Nagaosa2010,Shick2015,Kao2010}. Nevertheless, the prior FM FeRh measurements provide an upper bound estimate of such contaminating contributions to the charge current. Thus even in the case of a proximate magnetism FeRh may be regarded as a useful spin current detector.   

In conclusion, intrinsic Hall effects and their AFM-FM transitions in FeRh have been studied using a contemporary computational approach based on DFT/Wannier function representation. The inherent symmetries in the electronic structure and Berry phase have important signatures in the various Hall phenomena, including a sign change in the spin Hall conductivity. Thus FeRh with its $T$-controlled AFM-FM transition can be viewed as an alternative spin current detector. We hope our study will stimulate much needed experimental work in temperature dependent Hall effect measurements in FeRh. We believe that investigating thermally driven intrinsic Hall and Nernst effect changes in other materials exhibiting AFM-FM metamagnetic transitions, such as manganites, CeFe\textsubscript{2} or Mn\textsubscript{2}Sb alloys, will bring basic science forward and give new perspectives in spin-related applications.

\begin{acknowledgments}
Financial support from the US Department of Energy under Grant No. DE-FG02-06ER46297 is acknowledged. The use of the NIST Raritan HPC facilities is also acknowledged. P.R.-L. also acknowledges partial support from TerMic (Grant No. FIS2014-52486-R, Spanish Government), CONTRACT (Grant No. FIS2017-83709-R, Spanish Government) and from Juan de la Cierva - Incorporacion program (Ref: I JCI-2015-25315, Spanish Government).
\end{acknowledgments}

\bibliography{SHE_V10}

\begin{thebibliography}{45}
\bibitem [{com()}]{comment}%
  \BibitemOpen
  \href@noop {} {}\bibinfo {note} {This is unlike Rh-terminated surfaces, for
  which FM interface layers in the AFM FeRh phase have been
  observed.}\BibitemShut {Stop}%
\end{thebibliography}

\end{document}